\documentclass[draft]{agujournal2018}
\usepackage{apacite}
\usepackage{url} 
\journalname{Geophysical Research Letters}

\begin{document}

%
%


\title{Observations of Magnetic Reconnection in the Transition Region of Quasi-Parallel Shocks}

%
%




\authors{I. Gingell\affil{1}, S.J. Schwartz\affil{2,1}, J. P. Eastwood\affil{1}, J. E. Stawarz\affil{1}, J. L. Burch\affil{3}, R. E. Ergun\affil{2}, S. Fuselier\affil{3}, D. J. Gershman\affil{4}, B. L. Giles\affil{4}, Y. V. Khotyaintsev\affil{5}, B. Lavraud\affil{6}, P.-A. Lindqvist\affil{5}, W. R. Paterson\affil{4}, T. D. Phan\affil{7}, C. T. Russell\affil{8}, R. J. Strangeway\affil{8}, R. B. Torbert\affil{9}, F. Wilder\affil{2}}

\affiliation{1}{The Blackett Laboratory, Imperial College London, SW7 2AZ, United Kingdom}
\affiliation{2}{Laboratory for Atmospheric and Space Physics, University of Colorado, Boulder, Colorado 80303, USA}
\affiliation{3}{Southwest Research Institute, San Antonio, Texas 78238, USA}
\affiliation{4}{NASA, Goddard Space Flight Center, Greenbelt, Maryland 20771, USA}
\affiliation{5}{Swedish Institute of Space Physics (Uppsala), Uppsala, Sweden}
\affiliation{6}{Institut de Recherche en Astrophysique et Plan\'etologie, CNRS, UPS, CNES, Universit\'e de Toulouse, Toulouse, France}
\affiliation{7}{Space Science Laboratory,University of California, Berkeley, California, USA}
\affiliation{8}{University of California, Los Angeles, Los Angeles, California 90095, USA}
\affiliation{9}{University of New Hampshire, Durham, New Hampshire 03824, USA}





\correspondingauthor{Imogen Gingell}{i.gingell@imperial.ac.uk}




\begin{keypoints}
\item Reconnecting current sheets have been observed at a quasi-parallel bow shock.
\item The ion-scale current sheet exhibits only an electron jet and heating, with no ion response.
\item Consistent with hybrid simulations, reconnection relaxes complexity in the shock transition region.
\end{keypoints}

%
%


\begin{abstract}
Using observations of Earth's bow shock by the Magnetospheric Multiscale mission, we show for the first time that active magnetic reconnection is occurring at current sheets embedded within the quasi-parallel shock's transition layer. We observe an electron jet and heating but no ion response, suggesting we have observed an electron-only mode. The lack of ion response is consistent with simulations showing reconnection onset on sub-ion timescales. We also discuss the impact of electron heating in shocks via reconnection.
\end{abstract}

%
%

%


%
%
%
%

\section{Introduction}
\label{sec:intro}

Collisionless shocks are found in many astrophysical plasma environments, including planetary and stellar bow shocks, interplanetary shocks in the solar wind, and supernova remnants \citep{burgess_scholer_book_2015}. In order to reduce flows from super- to sub-sonic speeds, collisionless shocks must dissipate energy by particle processes, i.e. they are by necessity kinetic plasma structures. Understanding these microphysical processes is critical for understanding particle heating and acceleration \citep{auer1962,morse1972,gosling1985}.
The family of kinetic plasma processes responsible for energy dissipation is strongly dependent on shock parameters such as the Mach number, plasma beta, and the angle, $\theta_{Bn}$, between upstream magnetic field and shock normal \citep{burgess_scholer_book_2015}.

\begin{figure}
\includegraphics[width=\linewidth,clip,trim = 0 0 0 00]{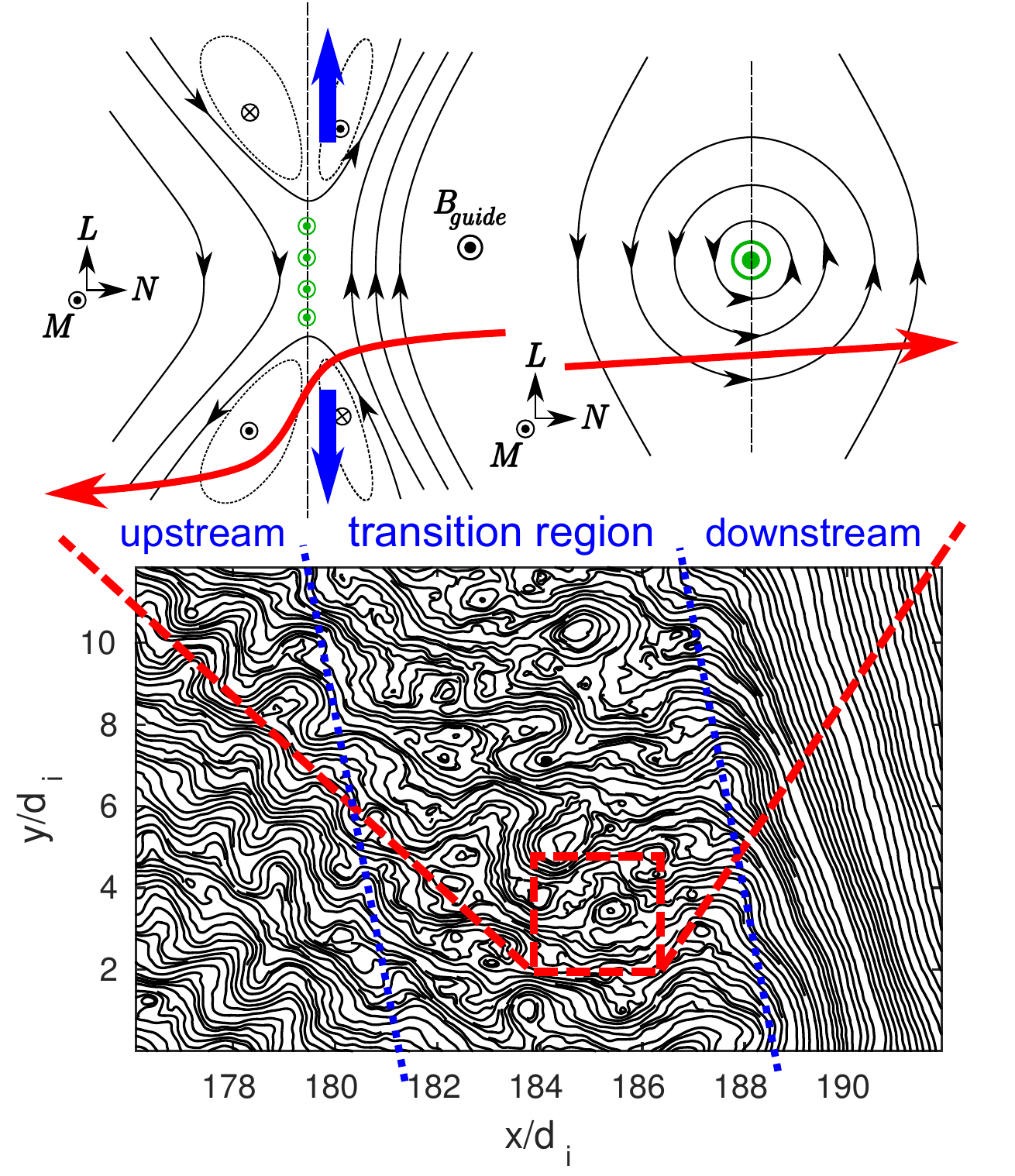}
\caption{Top: Schematics of the magnetic structure (black), out-flowing jet directions (blue) and current densities (green) for a reconnecting current sheet (left) and a flux rope (right). Red arrows depict the trajectories of MMS1 through the structures observed in Figures \ref{fig:jet}. Bottom: Snapshot of the magnetic field line structure of a hybrid simulation of a reforming quasi-parallel shock \citep{gingell2017}, demonstrating the appearance of current sheets and flux ropes within the transition region.}
\label{fig:scheme}
\end{figure}

In examining the non-stationary structure of quasi-parallel shocks ($\theta_{Bn} < 45^\circ$), recent simulations have shown that processes within the shock foot can generate current sheets and magnetic islands \citep{gingell2017}. The evolution of these regions is modulated by cyclic self-reformation. Reformation is a kinetic process driven by ions reflected from the shock ramp \citep{biskamp1972,hada2003,scholer2003a}, or by instabilities associated with whistler waves localised in the foot region \citep{scholer2007}, or by instabilities of the backstreaming ions in the foreshock \citep{burgess1989,kraussvarban1991,burgess1995}. Within the shock transition region, distinct from the magnetosheath downstream, magnetic islands merge to form larger scale structures that are convected towards the magnetopause. 
An example snapshot of one such simulation, revealing embedded current sheets and magnetic islands (flux ropes), is visible in Figure \ref{fig:scheme}. 
Within this model, self-reformation and other foot instabilities generate disordered or turbulent magnetic fluctuations close to the shock ramp. Decay of these disordered fluctuations may then occur via magnetic reconnection at current sheets and magnetic islands. These structures thus are closely associated with magnetic reconnection.

In this letter, we demonstrate for the first time that active magnetic reconnection is occurring in the transition region of Earth's quasi-parallel bow shock. We show that reconnecting current sheets are present within a disordered transition region close to the shock ramp, which is consistent with the appearance of these structures in recent hybrid and kinetic shock simulations \citep{gingell2017,karimabadi2014,matsumoto2015,bohdan2017}.
Magnetic reconnection, for which localised changes in magnetic topology result in rapid transfer of energy from fields to particles, has been observed in detail by Magnetospheric Multiscale (MMS) at Earth's magnetopause \citep{burch2016} and more recently in the turbulent magnetosheath \citep{phan2018}. In contrast to magnetosheath observations reported in \citep{phan2018}, structures discused here appear within seconds of crossing the bow shock, suggesting a close association with shock processes and a rapid evolution.
In the standard model, reconnection occurs within an electron-scale diffusion region \citep{vasyliunas1975,burch2016}, while at ion scales coupled ions are ejected from the diffusion region as bi-directional jets \citep{paschmann1979,phan2000,gosling2005}. Reconnection exhausts then extend to much larger scales.
In turbulent plasmas, magnetic reconnection is thought to play an important role in dissipation of energy at kinetic scales \citep{matthaeus1986,servidio2009,retino2007,sundqvist2007}.
Given the observations of electron heating detailed in this letter, we raise the question of how reconnection can contribute to shock energetics.
 
\section{A Reforming, Quasi-parallel Shock}
\label{sec:overview}

Here we discuss a crossing of Earth's bow shock by the four MMS spacecraft on 26 January 2017, 08:13:04 UTC. The mean spacecraft separation was 7km.
Electromagnetic field data are provided by the flux gate magnetometer (FGM) \citep{russell2016} and electric field double probe (EDP), both within the FIELDS suite \citep{torbert2016}. Particle data have been provided by the Fast Plasma Investigation (FPI) \citep{pollock2016}.
The sampling frequency is 128Hz for the FGM magnetic fields, and 8kHz for the EDP electric fields.
The full three-dimensional ion phase space is sampled by FPI every 0.15s, and the electron phase space is sampled every 0.03s. 

\begin{figure}
\includegraphics[width=\linewidth,clip,trim = 0 0 30 00]{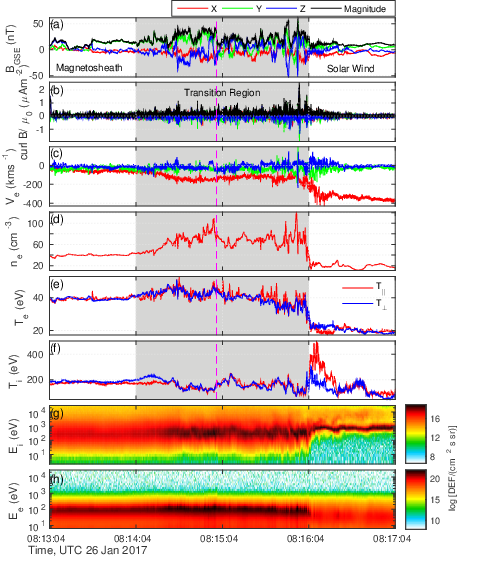}
\caption{Overview of the bow shock crossing observed by MMS1 on 26th January 2017, 08:13:04 UTC, in Geocentric Solar Equatorial (GSE) coordinates. From top to bottom: magnetic field, curl of the magnetic field, electron bulk velocity, electron number density, electron temperature, ion temperature, spectrograms of the differential energy flux for ions and electrons. A disordered transition region is evident for the period 08:14:04 to 08:16:04 UTC, shown in grey. The dashed magenta line shows the time of the event in Figure \ref{fig:jet}.}
\label{fig:overview}
\end{figure}

For the chosen event, the angle between the upstream magnetic field and shock normal is given by $\theta_{Bn} = 21^\circ$, the Alfv\'{e}nic Mach number of the upstream flow is $M_A = 3$, the fast magnetosonic Mach number is $M_{\mathrm{fast}} \approx 2$, and the upstream plasma beta is $\beta = 1.4$.
An overview of the event is shown for MMS1 in Figure \ref{fig:overview}. The magnetic field data in panel (a) demonstrates the presence of a transition region (highlighted in grey) between the relatively quiescent magnetosheath (before 08:14:04) and solar wind (after 08:16:04). Within this region the magnetic field is disordered, exhibiting multiple directional discontinuities. Using all four MMS spacecraft, we can use the curlometer method \citep{issibook_curlometer} to determine the barycentric current density, shown in panel (b). The high amplitude, narrow peaks within the current density (i.e. $\nabla \times B/\mu_0$) reveal several narrow current sheet-like structures with peak current densities on the order of $1\mu Am^{-2}$.
This transition region is associated with significant fluctuations of the electron velocity, and enhancements in the electron number density and temperatures. Although we also observe fluctuations in the ion temperatures, there is no enhancement across the full transition region.
We note that the change in field and plasma properties from the magnetosheath to the transition region at 08:14:04 may be in part associated with changes in the upstream plasma conditions rather than stationary shock structure.

In the solar wind, periodic reductions in the wind speed, visible at 08:16:20 and 08:16:40 in the ion differential energy flux and the bulk velocity $V_{eX}$ (panels (g) and (c)), suggest that, as with the simulation in Figure \ref{fig:scheme}, this shock may be undergoing cyclic self-reformation \citep{burgess1989} on a 20s timescale. 
Thus, this event is appropriate for evaluating the predictions of recent hybrid simulations of reforming, quasi-parallel shocks with respect to reconnection (see \citet{gingell2017} and Figure \ref{fig:scheme}).

\section{Current Sheets}
\label{sec:currentsheets}

For discussion of individual coherent structures, we introduce a new coordinate system derived by using a hybrid minimum variance analysis \citep{gosling2013,phan2018}.
The current sheet normal $N$ is determined using $\mathbf{B_1} \times \mathbf{B_2} /\left|\mathbf{B_1} \times \mathbf{B_2} \right|$, where $\mathbf{B_{1,2}}$ are the fields at the two edges of the current sheet. The $M$ direction, corresponding to the current carrying direction, is given by $\mathbf{M} = \mathbf{L'}\times \mathbf{N}$, where $\mathbf{L}'$ is the direction of the maximum variance of the magnetic field. Finally, $\mathbf{L} = \mathbf{N}\times\mathbf{M}$.

Although many magnetic directional discontinuities are visible within the transition region shaded in Figure \ref{fig:overview}, we must observe electron or ion jets in order to conclude that these current sheets are actively reconnecting. These jets, corresponding to outflow of plasma from an active reconnection site, are expected in the $L$-direction. Structures in bulk velocity may be unipolar if the spacecraft crosses only one jet, or bipolar if the spacecraft crosses both jets. A schematic of the magnetic field, current and jet directions is shown in the top left of Figure \ref{fig:scheme}.

\begin{figure}
\includegraphics[width=0.8\linewidth,clip,trim = 0 5 0 00]{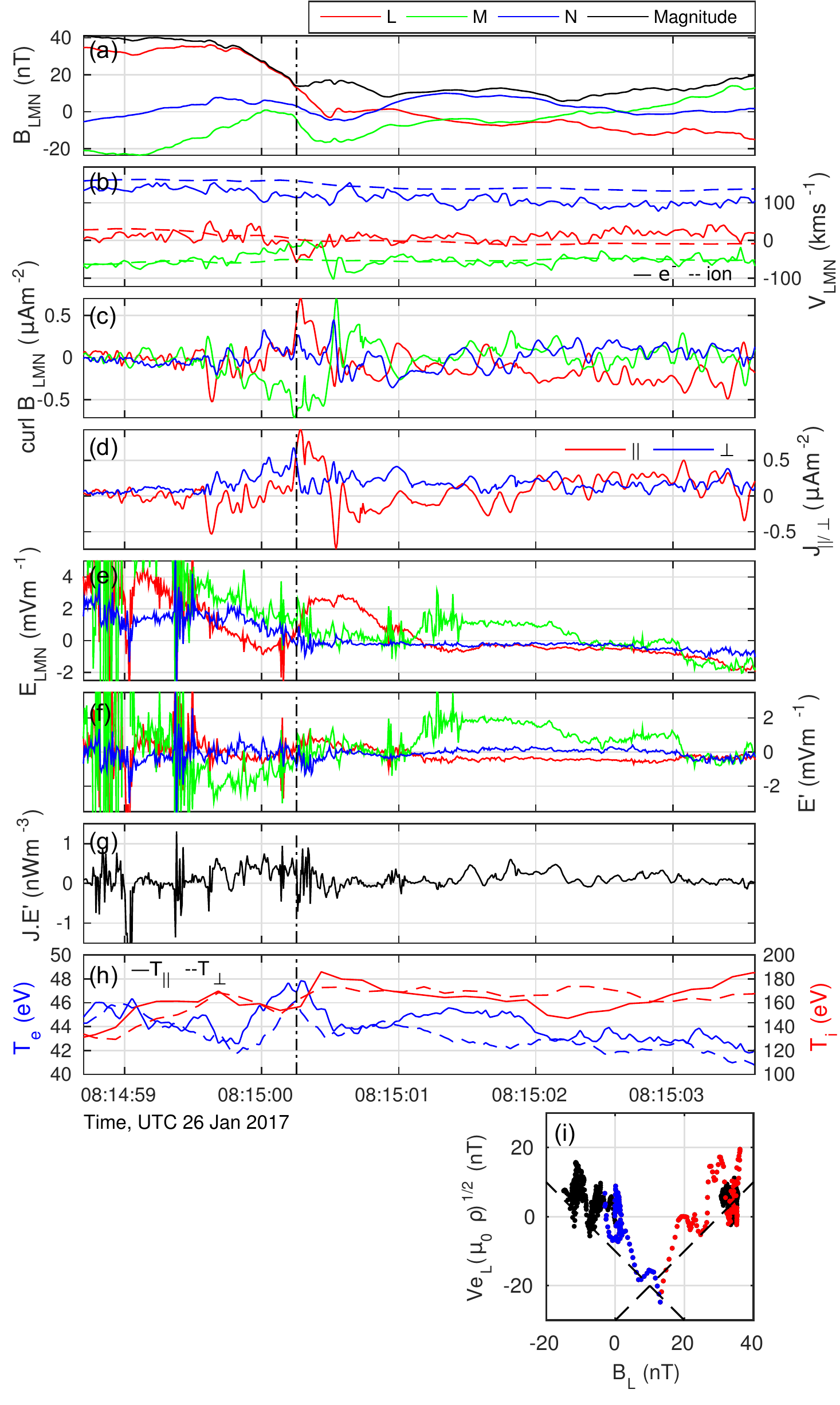}\\
\caption{Observation of a reconnecting current sheet within the transition region, presented in a minimum variance coordinate system, and in the spacecraft frame. From top to bottom: magnetic field, electron (solid) and ion (dashed) bulk velocity, curl of the magnetic field, current density parallel and perpendicular to the magnetic field, electric fields, $\mathbf{E}' = \mathbf{E}+\mathbf{V}_e\times \mathbf{B}$, heating measure $\mathbf{J}\cdot\mathbf{E}'$, and electron and ion temperatures scatter showing correlation of the $L$-component of the electron bulk velocity and magnetic field. The dashed vertical line is centered on the peak of the electron jet observed in $Ve_L$.}
\label{fig:jet}
\end{figure}

An example of a well-resolved current sheet with an electron jet is shown in Figure \ref{fig:jet}. Panel (a) shows the magnetic field components, demonstrating a change in sign of $B_L$ (red) over approximately 1s, a guide field with bipolar Hall fields in $B_M$ (green), and a reduction in field magnitude (black). The field magnitude is not symmetric across the current sheet; it transitions from 40nT to 20nT over 3s, with an intermediate plateau for 1.5s where $B_L \approx 0$. This is consistent with an asymmetric current sheet embedded within an inhomogeneous transition layer, with current and jet associated with the high-field side of the separatrix \citep{eastwood2013}. Under Taylor's hypothesis, using the normal component of the bulk velocity, this corresponds to a current sheet width of 3 ion inertial lengths. 

Panel (b), showing bulk velocities, reveals that the current in $V_M$ (green) is carried by the electrons. The ion bulk velocities (dashed lines) do not vary across the current sheet. 
The reconnection jet is visible in $Ve_L$ (red) as a deviation from the background velocity in the $-L$ direction, centered on the dashed vertical line. For a current sheet, a peak in the bulk velocity in the maximum variance $L$ direction is indicative of reconnection. It is important to note that no jet is visible in the ion bulk velocity. The peak electron velocity at the centre of the jet is approximately $1.2V_A$, for local parameters.
Given the directions of the magnetic field, current and electron jet, we can infer the trajectory of spacecraft through an idealised reconnection site. This trajectory is shown with a red arrow in the top left of Figure \ref{fig:scheme}.
We note that all four MMS spacecraft observe similar features, suggesting all four cross the current sheet on the same side of the diffusion region.

The appearance of a reconnecting electron jet is further supported by the correlation between $Ve_L$ and $B_L$. A scatter plot is shown inset in Figure \ref{fig:jet}. The jet is Alfv\'{e}nic, lying principally along the Walen slopes $B_L \propto \pm Ve_L(\mu_0\rho)^{1/2}$ (dashed lines), positively correlated approaching the electron jet (red points), and anti-correlated on passing the electron jet (blue points).

The electron jet is coincident with peaks in the perpendicular and parallel electron temperatures, corresponding to a 3eV increase. The mean electron temperature increase across the current-carrying region (08:14:59.5 to 08:15:01) is 0.5eV. However, as with the bulk velocities, ion temperatures do not show similar peaks. This further suggests that ions are not coupled to reconnection processes for this current sheet, despite the fact that the current sheet width is on the order of the ion inertial length. 
Another measure of heating, $\mathbf{J}\cdot\mathbf{E}'$, where $\mathbf{E}' = \mathbf{E}+\mathbf{v}_e\times \mathbf{B}$, is shown in panel (g). This corresponds to the exchange of energy between particles and fields in the particle rest frame.
Such a feature may be visible for 0.5s before the peak velocity of the electron jet. However, given the fluctuations of similar magnitude in the preceding second of the interval, it is unclear whether this feature is linked to ongoing heating driven by reconnection.

The preceding analysis demonstrates that reconnection occurs within the transition region of a quasi-parallel shock. Similarly to recent observations of magnetosheath reconnection \citep{phan2018}, the outflow jet and particle heating appear limited to electrons. However, in this example, the current sheet width is larger; on the order of the ion inertial length. Given the lack of ion response, this suggests that this feature is relatively young, on the order of the ion gyro-period or less, and may have just formed. This is supported by recent hybrid simulations, which suggest that reformation-driven generation of current sheets occurs on timescales faster than the ion gyro-period \citep{gingell2017}. It may be that an ion jet exists further from the reconnection site than the spacecraft trajectories pass. Although we do not observe clear ion jets for any other potential reconnection event associated with this shock, it may be that ion jets embedded within the turbulent structure of the transition region exhibit unexpected orientations.

\section{Conclusion}
\label{sec:conclusion}

Using observations of Earth's bow shock by MMS, we have demonstrated that reconnecting current sheets are present in the transition region of quasi-parallel shocks. 
Several reconnection jets have been observed within the shock shown in Figure \ref{fig:overview}, the clearest of which is shown in Figure \ref{fig:jet}. A further example of a current sheet, discovered within the transition region of another bow shock crossing observed by MMS on 31st December 2016, 06:06:24 UTC, is shown in the supporting material.
The observation of current sheets is consistent with the structure of the transition region reported in hybrid simulations by \citet{gingell2017}.
Magnetic reconnection may therefore play an important role in the energetics of collisionless shocks.

Observations of the magnetopause suggest that 1.7\% of the available inflow magnetic energy is transferred to the electrons during reconnection \citep{phan2013}, i.e. $\Delta T_e = 0.017 m_i V_{AL,\mathrm{inflow}}$. 
For the asymmetric current sheet detailed in Figure \ref{fig:jet}, this is given by $\Delta T_e =  0.2eV$. 
This is consistent with a mean electron temperature increase of 0.5eV across the current sheet. However, we note that reconnection at the shock appears to partition energy differently to magnetopause reconnection, favouring the electrons.
The total heating across the transition region can be seen in panels (e) and (f) of Figure \ref{fig:overview}. We find that the electron temperature rises from 20eV to 33eV in the ramp, and continues to rise another 7eV within the transition region. 
Thus, 35\% of the total shock electron heating occurs in the transition region. We note that no similar trend is visible in the ion temperatures, suggesting again that dissipative processes in this region affect only electrons.
We can estimate the ability of reconnection to provide the observed 7eV heating by considering the magnetic energy of the fluctuations per electron, $E_f = \left<(\delta B)^2\right>/(2\mu_0 n_e)$, where $\delta B = \left|\mathbf{B} - \left<\mathbf{B}\right>\right|$. For the transition region shown in Figure 2, $E_f = 20 eV$ per electron, while in the magnetosheath $E_f = 10eV$. 
A 10eV dissipation is consistent with the observed 7eV electron temperature increase across the transition region. However, further work is required to establish the balance between reconnection and other dissipative processes in accounting for this temperature change.

Mechanisms for electron heating are strongly dependent on shock parameters such as the Mach number, $\theta_{Bn}$, and plasma betas \citep{ghavamian2013}. At supernova remnants, heating can be driven by waves excited by shock reflected ions or streaming cosmic rays, via the lower hybrid drift instability \citep{ghavamian2007} or the Buneman instability (for $M_A > 50$) \citep{cargill1988}. Within the solar wind, heating may be driven by a modified two-stream instability or electron cyclotron drift instability \citep{umeda2012,matsukiyo2010}, or simply by the cross shock potential \citep{lefebvre2007}. However, these mechanisms are most efficient for quasi-perpendicular shocks. Thus, the observation of reconnection-driven heating at a quasi-parallel shock represents an important development in the characterisation of energy partition at shocks in both astrophysical and space plasmas.

The reconnection event featured in this paper represents a regime in which the current and reconnection outflows are associated only with electrons, similar to the magnetosheath event reported by \citep{phan2018}. However, in this case the scale lengths are on the order of the ion inertial scale. This suggests that these current sheets are not generated by a turbulent cascade to electron scales. 
The lack of ion response, however, is consistent with a rapid onset time. Given the proximity of the current sheets to the shock ramp, and the timescale for cyclic reformation for similar shocks \citep{gingell2017}, the observed reconnection site may be younger than an ion gyro-period.
In simulations \citep{gingell2017}, rapid onset reconnection is driven at ion scales by instabilities in the foreshock and foot, generating magnetic islands in the transition region. 
These instabilities, modulated by cyclic reformation, may generate a range of scales simultaneously, rather than by ongoing cascade as expected in the magnetosheath turbulence.
These structures then coalesce via secondary reconnection as they convect downstream, relaxing the magnetic field.

These observations support the need for more detailed simulations of reconnection at shocks, and observational surveys across all parameter regimes. 
This will allow us to asses the broader impact of reconnection on heating and particle acceleration at collisionless shocks, explore the evolution of these structures as they convect downstream, and determine how reconnection properties at coherent, rapidly-driven thin boundaries differ from models of reconnection operating elsewhere in the magnetosphere and heliosphere.

\acknowledgments
This work was supported by the UK Science and Technology Facilities Council (STFC) grant ST/N000692/1. Data used in this research is publicly available at the MMS Science Data Center at the Laboratory for Atmospheric and Space Physics (LASP) hosted by the University of Colorado, Boulder (https://lasp.colorado.edu/mms/sdc/public/). Work at IRAP is supported by CNRS and CNES.


%
%




\end{document}